# The SLcam: A full-field energy dispersive X-ray camera

A. Bjeoumikhov [a], G. Buzanich [b], N. Langhoff [a], I. Ordavo [c,d], M. Radtke [b], U. Reinholz [b], H. Riesemeier [b], O. Scharf [e], H. Soltau [c], and R. Wedell [e]

[a] *IFG Institute for Scientific Instruments GmbH.,*
  *12489 Berlin, Germany*

[b] *BAM Federal Institute for Materials Research and Testing,*
  *12489 Berlin, Germany*

[c] *PNSensor GmbH.,*
  *80803 Munich, Germany*

[d] *PNDetector GmbH.,*
  *81735 Munich, Germany*

[e] *IAP Institute for applied Photonics e.V.,*
  *12489 Berlin, Germany*
  *E-mail*: scharf@ifg-adlershof.de

ABSTRACT: The color X-ray camera (SLcam®) is a full-field single photon imager. As stand-alone camera, it is applicable for energy and space-resolved X-ray detection measurements. The exchangeable poly-capillary optics in front of a beryllium entrance window conducts X-ray photons from the probe to distinguished energy dispersive pixels on a pnCCD. The dedicated software enables the acquisition and the online processing of the spectral data for all 69696 pixels, leading to a real-time visualization of the element distribution in a sample. No scanning system is employed. A first elemental composition image of the sample is visible within minutes while statistics is improving in the course of time. Straight poly-capillary optics allows for 1:1 imaging with a space resolution of 50 μm and no limited depth of sharpness, ideal to map uneven objects. Using conically shaped optics, a magnification of 6 times was achieved with a space resolution of 10 μm.

We present a measurement with a laboratory source showing the camera capability to perform fast full-field X-ray Fluorescence (FF-XRF) imaging with an easy, portable and modular setup.

## Contents

## 1. Introduction

Spatially resolved X-ray fluorescence micro analysis (µXRF) is until now possible only by using a focused X-ray beam that defines the spot (pixel) size. The spectrum for each spot is analyzed by a zero-dimensional (non spatial resolving) detector. For a chemical mapping of a whole object each point is measured separately by moving the sample step by step. If the sample is uneven each point needs to be refocused leading to a time consuming procedure. An actual development of X-ray detection technology and X-ray poly - capillary optics has overcome these limitations. By combining a pnCCD detector chip with a full field of view of (12.7 x 12.7) mm$^2$ with special designed poly-capillary optics, one compact detection device, the SLcam® was developed[1,2]. A unique advantage of this so called X-ray color camera is the possibility to perform direct high resolution full-field energy-resolved X-ray imaging of the sample surface. As no scanning system needs to be employed, the stand-alone camera can replace conventional X-ray detectors to extend the experiments to full field analysis. This advantage of the camera is a big step forward for efficient chemical characterization of materials and will open completely new possibilities to analyze all kinds of samples in fields of e.g. materials sciences, medicine, biology, archaeology, geology or recycling. Several experiments have been performed, focused mainly on the use of synchrotron radiation [3,4,5].
The optics of the camera is exchangeable. The 1:1 optics has an unlimited depth sharpness as no focused beam is used – ideal to map uneven objects. Using different types of X-ray poly capillary optics with varying magnification it is also possible to increase or decrease the local resolution or, vice versa, the area that is measured[6,7]. Details regarding the pnCCD of the camera are given in [1,2].

## 2. Experimental Setup

The camera head is vacuum sealed and has a 50 µm Be entrance window. The pnCCD is 6 mm away from the window and is thermally coupled to a two stage Peltier element. At a working temperature of -26°C the energy resolution is 152 eV for Mn Kα[1]. The principal experimental setup is shown in Figure 1. The sample is placed under 45° towards source and SLcam®. The distance to the source is optimized to illuminate the full field of view on the sample. The distance sample optics is as close as the space available from the stage setup. A simple hand operated xyz-stage is used as sample holder.

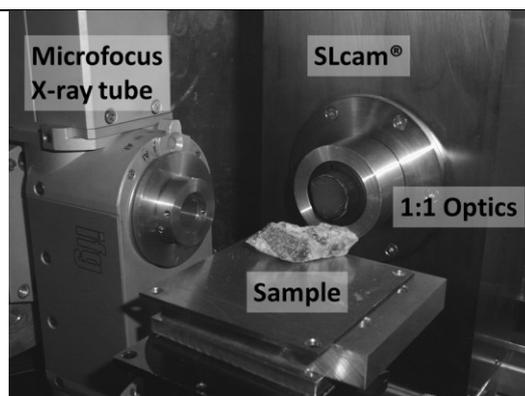

Figure 1: Experimental laboratory setup at the Institute for Scientific Instruments (IFG) with Rh microfocus X-ray tube, SLcam® with 1:1 optics and sample.

The chip is fully depleted and thus sensitive over its whole depth of 450 µm. Due to the 50 µm Be-window the lower detection energy limit is around 2 keV. Up to 10 keV, the chip has quantum efficiency of almost 100% and at 20 keV the efficiency is still above 35 %. The optics used has a transparency over 75%. For photons above 20 keV, the glass itself becomes transparent, setting the high energy imaging limit to photons with energy around 20 keV. The camera is operated by dedicated fully programmable driving and acquisition software. The software enables to obtain online 2D color maps of the distribution of each element selected by summing the number of counts in energy regions for the elemental fluorescence lines (regions of interest (ROIs)), a global 2D color map of the distribution of all the elements selected of the sample (overlay image) and sum spectra with 2000 channels of selected regions. The numbers of counts within the ROI regions are mapped linearly to a color interval from dark to a defined color. Counts below a user set lower cut off are mapped to transparent (no color set). Counts above a user set upper cut off are mapped to the defined color. The lower and upper cut off are indicated next to the color bar in the map image. The name of the line, the ROI width and measurement time are indicated, too. The overlay images are built by taking for each pixel the color value of the ROI map with the highest value of counts times a factor. The factor was adapted in a way to enhance and bring important elemental features to the front. The channel diameter of the optics is with 20 µm smaller than the pnCCD pixel size. Thus the spatial resolution is given by the 48 µm of the CCD pixel.

## 3. Application

The easy integration and full field measurement with the camera is demonstrated on a geological Titanit sample (kindly provided by GEOMONTAN, Gesellschaft für Geologie und Bergbau mbH & Co. KG Sachsen). It was analyzed using a low power 30 Watt Rh micro focus tube placed 5 cm from the sample to illuminate it completely. The sample – camera optics distance was 1 cm. The photo and composition image obtained by the camera are indicated on the left of Figure 2. The single element maps for Ca, Ti and Fe are given on top. The sum spectra is presented below with the element fluorescence lines and regions of interest indicated according to their assigned color. The total measuring time was 2h with a count rate of 480 cps and the image was visible already in less than one minute. The sample was not prepared (e.g. sanded) and placed as it was in front of the camera.

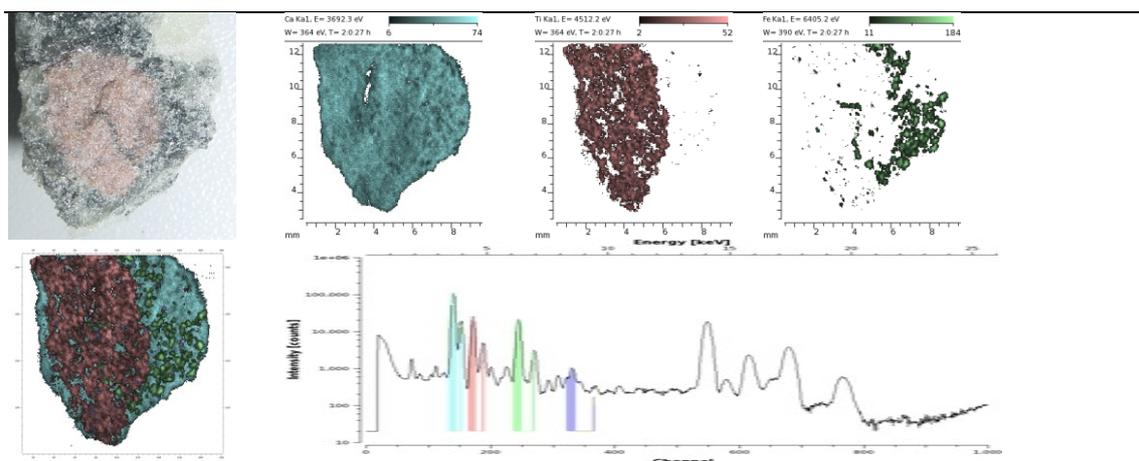

Figure 2: Example measurement of a Titanit geological sample (kindly provided by Dr. Mucke, GEOMONTAN). On the Left: Light optical image and composite SLcam® image. Middle: Single element maps for Ca, Ti and Fe. Bottom: Sum spectra with indicated regions of interest. Resolution 50 µm for 1 cm² field of view, measurement time 2h (480cps).

## 4. Conclusions

A new device for simultaneous spatial and energy-resolved X-ray imaging has been presented. The SLcam® is a stand-alone device without moving parts that is easy to integrate into existing setups because only the standard detector needs to be replaced. As no focusing beam is employed and the 1:1 poly - capillary optics has an unlimited depth of sharpness, the imaging of uneven sample surfaces becomes possible without time consuming refocusing. Dedicated software allows the visualization of the color coded element distribution of the samples in real time. Further work will use a window less camera to image low energy lines and to implement new reconstruction algorithms for position resolution well below the physical pixel size.

## References

[1] Scharf, O. et al., *Compact pnCCD-Based X-ray Camera with High Spatial and Energy Resolution: A Color X-ray Camera*, 2011 Anal. Chem., Vol. 83(7), pp. 2532 – 2538.

[2] Ordavo, I., et al., *A new pnCCD-based color X-ray camera for fast spatial and energy-resolved measurements*, 2011 Nucl. Instrum. Methods Phys. Res. Sec. A 654(1) pp. 250 – 257.

[3] Kühn, A. et al., *Pushing the limits for fast spatially resolved elemental distribution patterns*, 2011 J. Anal. At. Spectrom., Vol. 26, pp. 1986 – 1989.

[4] Donges, J. et al., *Energy dispersive X-ray diffraction imaging*, 2011 MECA SENS VI

[5] A. Guilherme et al., *Synchrotron micro-XRF with Compound Refractive Lenses (CRLs) for tracing key elements on Portuguese glazed ceramics*, 2012 J. Anal. At. Spectrom., Vol. 27, pp. 966 – 974.

[6] Arkadiev, A. et al., *X-ray focusing by polycapillary arrays*, 1995 Proc. SPIE. 2515, pp. 514 – 525.

[7] Bzhaumikhov, A. et al., *Polycapillary conic collimator for micro XRF*, 1998 Proc. SPIE 3444, pp. 430 – 435.